\documentclass[aps,pra,reprint,twocolumn,superscriptaddress,showpacs]{revtex4-1}

\usepackage{graphicx,dcolumn,hyperref,braket,amsmath,amssymb} 
\usepackage{epstopdf,color}
\usepackage{natbib}

\begin{document}

%\title[A quantum version of the negative binomial distribution]{A quantum version of the negative binomial distribution as a universal photon distribution for a cavity containing two atoms}
%\title[Quantum signatures of collective behaviour]{A quantum version of the negative binomial distribution and further quantum signatures of collective behaviour}
%\title[Nonclassical correlations of two atoms coupled to a cavity]{Nonclassical correlations of two atoms coupled to a cavity: A quantum version of the negative binomial distribution}
\title{Phase Control of the Quantum Statistics of Collective Emission}

\author{M.-O. Pleinert}
%\thanks{Correspondence and requests for materials should be addressed to M.-O.P. (marc.pleinert@fau.de)}
\affiliation{Institut f\"{u}r Optik, Information und Photonik, \\Friedrich-Alexander-Universit\"{a}t Erlangen-N\"{u}rnberg (FAU), 91058 Erlangen, Germany}
\affiliation{Erlangen Graduate School in Advanced Optical Technologies (SAOT), Friedrich-Alexander-Universit\"{a}t Erlangen-N\"{u}rnberg (FAU), 91052 Erlangen, Germany}
\author{J. von Zanthier}
\affiliation{Institut f\"{u}r Optik, Information und Photonik, \\Friedrich-Alexander-Universit\"{a}t Erlangen-N\"{u}rnberg (FAU), 91058 Erlangen, Germany}
\affiliation{Erlangen Graduate School in Advanced Optical Technologies (SAOT), Friedrich-Alexander-Universit\"{a}t Erlangen-N\"{u}rnberg (FAU), 91052 Erlangen, Germany}
\author{G. S. Agarwal}
\affiliation{Institute for Quantum Science and Engineering, Department of Biological and Agricultural Engineering and Department of Physics and Astronomy, Texas A\&M University, College Station, Texas 77843, USA}

\date{\today}

%\pacs{42.50.Pq, 42.50.Nn, 37.30.+i}

\begin{abstract}
We report nonclassical aspects of the collective behaviour of two atoms in a cavity by investigating the photon statistics and photon distribution in a very broad domain of parameters. Starting with the dynamics of two atoms radiating in phase into the cavity, we study the photon statistics for arbitrary interatomic phases as revealed by the second-order intensity correlation function at zero time $g^{(2)}(0)$ and the Mandel $Q$ parameter.
We find that the light field can be tuned from antibunched to \mbox{(super-)}bunched as well as nonclassical to classical behaviour by merely modifying the atomic position.
The highest nonclassicality in the sense of the smallest $Q$ parameter is found when spontaneous emission, cavity decay, coherent pumping, and atom-cavity coupling are of comparable magnitude. 
We introduce a quantum version of the negative binomial distribution with its parameters directly related to $Q$ and $g^{(2)}(0)$ and discuss its range of applicability. We also examine the Klyshko parameter which highlights the nonclassicality of the photon distribution.
\end{abstract}

%\pacs{42.50.Pq, 42.50.Nn, 37.30.+i}
%\noindent{\it Keywords}: Dicke superradiance, atoms in a cavity, photon statistics, negative binomial distribution
%\submitto{\NJP}

\maketitle

\section{Introduction}

Nonclassical properties of light, i.e., effects and phenomena, which cannot be explained by classical optics, have been studied extensively in recent years. The light emitted from a few atoms confined to a cavity, for instance, can exhibit remarkable quantum effects like antibunching or squeezing \cite{Paul:1986,Loudon:1987,Walls:2008,Agarwal:2012}. 
%Nonclassical light, however, is not only intriguing for fundamental questions in quantum physics, but also possesses applications in superresolving imaging, e.g., via NOON-states \cite{Boto:2000,Dowling:2008,Afek:2010} or in quantum information processing \cite{Nielsen:2000,Kimble:2008,Specht:2011}. 

Measures for the nonclassicality of light fields have been transferred from quantum mechanics like phase-space quasi-probability distributions such as the Wigner function $W(q,p)$, but also new ones were introduced in the context of the quanta of light like the Mandel $Q$ parameter \cite{Mandel:1979} or the squeezing parameter $S$ \cite{Yuen:1976,Caves:1985,Schumaker:1985}. The latter indicators are based on investigations of the fluctuations in the number of photons or in the field quadratures, respectively. Nowadays, a great number of such nonclassicality measures \cite{Agarwal:1992a,Richter:2002,Shchukin:2005,Rivas:2009} and generalizations to multimode fields \cite{Miranowicz:2010} exist. Yet, most of these nonclassicality measures are sufficient but not necessary. Richter and Vogel formulated a necessary and sufficient hierarchy of conditions for nonclassicality utilizing the Bochner theorem from probability theory \cite{Richter:2002}. Till today, however, no simple general criteria that is sufficient and necessary has been found \cite{Strekalov:2017}. 
In order to focus on the underlying physics, we will narrow our investigations of nonclassicality to quantities that are based on comprehensible physical principles, namely sub- or super-Poissonian statistics of the photon distribution and (anti-)bunching of photons. These are the Mandel $Q$ parameter and Glauber's normalized second-order correlation function at zero time $g^{(2)}(0)$ \cite{Glauber:1963}, respectively. 

In this article, we study the nonclassicality of light fields due to cooperative atomic emission. When studying collective effects, configurations consisting of just two emitters are particularly interesting as they serve as the most elementary system to study cooperative effects. Moreover, to keep the discussion clear, it is desirable to focus on a single-mode of the electromagnetic field. Very recently, three groups reported the experimental achievement of such an ideal system, namely two precisely steerable atoms \cite{Reimann:2015,Neuzner:2016} or ions \cite{Casabone:2015} coupled to a single-mode cavity.
This setup constitutes a natural platform to study quantum features of collective light emission as a function of the various parameters of the system.

Most theoretical studies of atom-cavity interactions assume a symmetric coupling. These include the original study by Tavis and Cummings \cite{Tavis:1968} and most of the recent investigations of a few atoms coupled to a cavity (e.g. \cite{Temnov:2009,Meiser:2010,Meiser:2010a}). 
Only few theoretical articles have been devoted to asymmetrical coupling configurations \cite{Zippilli:2004,Zippilli:2004a,Fernandez-Vidal:2007,Pleinert:2017}. In these studies, however, merely the rate of output photons \cite{Zippilli:2004,Zippilli:2004a,Pleinert:2017}, or very specifically for the dispersive cavity, the squeezing of the output spectrum \cite{Fernandez-Vidal:2007} have been discussed.
%The recent experimental realizations also mainly discuss the implications of an asymmetrical coupling on the mean photon rate \cite{Reimann:2015,Casabone:2015} or the }
There is, however, no investigation exploring the photon statistics of the specific systems in a very broad regime of parameters, which is the focus of the present study.
In contrast to the previously mentioned theoretical considerations, we neither make use of any approximations like adiabatic separation \cite{Temnov:2009,Meiser:2010a,Meiser:2010} nor limit ourselves to the strong coupling domain \cite{Zippilli:2004,Zippilli:2004a} nor describe the dynamics in the limit of large detuning \cite{Fernandez-Vidal:2007}.

%but explore the dynamical behavior in a very broad regime of parameters recovering the previous findings in the appropriate limits. 

We focus our study on the implications of the interatomic phase induced by an asymmetric coupling.
This phase control is, for instance, used in the classical case to enhance the mean signal, i.e., antennas oscillating in phase, and in the quantum case, a previous study showed that two atoms radiating out of phase are surprisingly able to produce a larger mean signal than the corresponding in-phase radiating atoms \cite{Pleinert:2017}. Here, we examine the phase control of quantum statistical aspects.
The specific findings of this article are (a) tuning from antibunching to superbunching of the emitted light by crossing over from an in- to out-of-phase configuration of the atoms, (b) the most prominent nonclassical behaviour when spontaneous emission, cavity decay, coherent pumping, and atom-cavity coupling are of comparable magnitude, (c) a photon number distribution for the collective radiation described by a quantum version of the negative binomial distribution with its parameters directly related to the two nonclassicality measures $g^{(2)}(0)$ and $Q$, and (d) nonclassicality of the photon number distribution as revealed by Klyshko's criterion.
%

%Although we study the cavity system in the context of atoms, the results should be applicable to any kind of two-level systems like ions \cite{Casabone:2015,DeVoe:1996,Stute:2012}, superconducting qubits \cite{Wallraff:2004,Fink:2009,Mlynek:2014} and quantum dots \cite{Hennessy:2007,Faraon:2008,Leymann:2015}. 

\section{System}

\begin{figure}
	\centering \includegraphics[width=0.9\columnwidth]{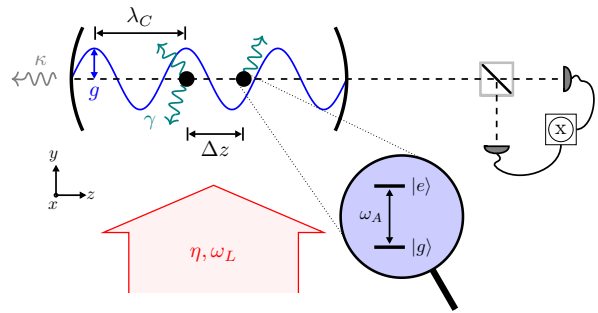}
	\caption{\label{fig:cavity_setup} Sketch of the investigated system. Two two-level atoms (with upper level $\ket{e}$, lower level $\ket{g}$ and transition frequency $\omega_A$) are driven by a coherent laser at frequency $\omega_L$ with Rabi frequency $\eta$ and couple to a single mode of a cavity (maximum coupling strength $g$) with frequency $\omega_C=2\pi c/\lambda_C$. Possible dissipative processes are spontaneous emission ($\gamma$) and cavity decay ($\kappa$). Along the cavity axis, intensity as well as intensity correlations can be measured by a Hanbury Brown and Twiss setup consisting of a beam splitter and two detectors.}
\end{figure}

A schematic drawing of the investigated system is shown in Fig. \ref{fig:cavity_setup}. The atoms - modeled as two-level systems with upper level $\ket{e}$ and lower level $\ket{g}$ at transition frequency $\omega_A$ - couple to a single mode $\omega_C$ of the cavity. 
An external laser field oriented perpendicular to the cavity axis coherently drives the atoms at frequency $\omega_L$. 
The emission of the system can, for instance, be measured by a Hanbury Brown and Twiss (HBT)-like setup \cite{Hanbury-Brown:1956} consisting of a beam splitter dividing the radiated intensity into two modes enabling one to record the intensity as well as second-order intensity correlations. For correlations of higher order, an accordingly extended setup has to be used. As the laser operates perpendicular to the cavity axis, it is guaranteed that values measured at the detectors are proportional to their corresponding intracavity value.
The atomic Hilbert space is given by the appropriate tensor product of the single atom-space, i.e., $\mathcal{H}\equiv (\mathbb{C}^2)^{\otimes 2}$, which is spanned by the factorized states $\ket{gg},\ket{eg},\ket{ge},\ket{ee}$.
The atoms can be fully characterized by spin-half operators, which for the $i$th atom read
\begin{eqnarray}
S_i^+ = \ket{e}_i\bra{g}_i \, , \quad
S_i^- = \ket{g}_i\bra{e}_i \, , \quad
S_i^z = \ket{e}_i\bra{e}_i \, .
\end{eqnarray}
The single intracavity mode, on the other hand, is described by the bosonic annihilation and creation operator $a$ and $a^\dagger$. 

\subsection{Hamiltonian and master equation treatment}

In an interaction frame rotating at the laser frequency $\omega_L$, the closed dynamics of atom and cavity are governed by the following Hamiltonian 
\begin{eqnarray}\label{eq:Hamiltonian}
H &=& H_0 + H_I + H_L \\
 &=& \hbar \Delta \sum_{i=1,2} S^z_i + \hbar \delta a^\dagger a  \nonumber \\
&& + \hbar \sum_{i=1,2} g_i \left( S_i^+ a + S_i^- a^\dagger \right) \nonumber \\
&& + \hbar \eta \sum_{i=1,2} \left( S_i^+ + S_i^-  \right) \, , \nonumber
\end{eqnarray}
which describes the atoms as well as their coupling to the cavity and the external laser. The three parts of the Hamiltonian describe the independent system of two atoms and a single-mode of the electromagnetic field ($H_0$), the interaction of the atoms and the cavity ($H_I$), as well as the driving of the atoms by the laser ($H_L$).

In the description of the independent system, $\Delta=\omega_A-\omega_L$ and $\delta=\omega_C-\omega_L$ constitute the atom-laser and cavity-laser detuning. The atom-cavity interaction is modelled by the Tavis-Cummings Hamiltonian \cite{Tavis:1968} which utilizes the dipole approximation and the rotating wave approximation. Unlike most previous studies, we do not assume the atoms to be symmetrically coupled to the cavity mode, but take into account that the atom-cavity-interaction crucially depends on the precise position of the atoms within the cavity. The coupling strength of the $i$th atom can thus be written as a function of its position $z_i$ and is given by $g_i=g\cos (2\pi z_i / \lambda_C)$, with $g$ being the maximum coupling constant. 
In this paper, we fix one atom at an antinode of the cavity field, while we vary the position of the other atom along the cavity axis. The former atom thus couples maximally to the cavity, while the coupling strength of the latter atom depends on the resulting interatomic distance $\Delta z=z_2-z_1$, inducing a phase shift $\phi_z=2\pi\Delta z / \lambda_C$ between the radiation emitted by the two atoms \footnote{Neutral atoms are stored in standing-wave traps leading to discrete atomic positions, i.e., multiples of the lattice parameter $\lambda_{trap}/2$. Here, we assume $\lambda_{trap} \ll \lambda_C$ allowing for a continuous scanning of the atom position and thus for a continuous phase variation. Note that in ion traps the interionic distance results from the interplay of repelling Coulomb interaction and attracting trap potential. By varying the trap potential via the trap frequency, the interionic distance can be continuously scanned.}.
The resulting atom-cavity-couplings can then be written as
\begin{subequations}
\begin{eqnarray}
g_1 &=& g \, , \\
g_2 &=& g \cos (\phi_z) \, .
\end{eqnarray}
\end{subequations}
%We note that we can choose $\phi_z$ modulo ${2\pi}$ such that direct atom-atom interactions as in \cite{Goldstein:1997} can be avoided by using separations of the atoms much larger than the cavity wavelength $\lambda_C$.
%
The coherent pumping of the atoms by an external laser is characterized by the Hamiltonian $H_L$. For a fixed atomic transition dipole moment $d_{eg}$, the Rabi frequency $\eta=d_{eg}\mathcal{E}/\hbar$ can be used to indicate the strength of the coherent pump field $\mathcal{E}$. We assume a homogeneous driving of the atoms, which is, for instance, the case, when the interatomic displacements in $y$-direction are negligible. Varying pump rates due to spatial variation of the laser phase could also be absorbed into modified coupling constants of the atoms.

Due to the non-neglectable dissipative processes, which are predominantly cavity decay, i.e., the leakage of intracavity photons from the cavity at a rate $\kappa$, and spontaneous emission by the atoms, i.e., the emission of photons into the side-modes at a rate $\gamma$, the dynamics of the entire system are to be described by the following master equation \cite{Agarwal:2012}
\begin{equation}\label{eq:master}
\frac{\partial}{\partial t} \rho =  -\frac{i}{\hbar} \left[H, \rho \right] + \mathcal{L}_\gamma \rho + \mathcal{L}_\kappa \rho \, ,
\end{equation}
where $\rho$ is the density operator describing the atom-cavity system. 
The dissipative processes, i.e., the impact by the environment, are included in the treatment by the Liouvillian superoperators $\mathcal{L}_\gamma$ and $\mathcal{L}_\kappa$. In particular, the sideway radiation is taken into account by the term 
\begin{equation}
\mathcal{L}_\gamma \rho = \frac{\gamma}{2} \sum_{i=1,2} \left( 2 S_i^- \rho S_i^+ - S_i^+S_i^- \rho - \rho S_i^+ S_i^- \right) \, ,
\end{equation}
while the leakage of photons is considered by the Liouvillian 
\begin{equation}
\mathcal{L}_\kappa \rho =  \frac{\kappa}{2} \left( 2a\rho a^\dagger -a^\dagger a \rho - \rho a^\dagger a \right) \, .
\end{equation}
Note that in this paper, we neglect dephasing effects, which are marginal for atoms and ions, but would become relevant in solid-state systems, like quantum dots.

\subsection{Energy levels and occurring transitions}

\begin{figure}
	\centering \includegraphics[width=\columnwidth]{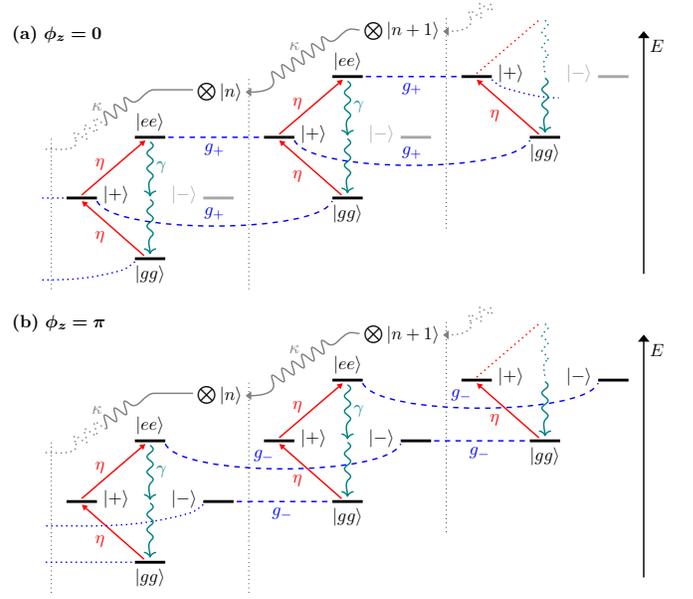}
	\caption{\label{fig:transitions} Overview of the energy levels and occurring transitions. For different manifolds of cavity states $\ket{n}$, the four atomic basis states are drawn. The lowest energy state shown thus corresponds to $\ket{gg}\otimes\ket{n}=\ket{gg,n}$. From this state, the external laser ($\eta$) only pumps the symmetric states, i.e., symmetric Dicke state $\ket{+}$ and doubly excited state $\ket{ee}$. Spontaneous emission ($\gamma$) deexcites the atoms within a specific manifold of cavity states, while cavity decay ($\kappa$) does not alter the atomic state, but shift the states between different cavity state-manifolds. The atom-cavity-coupling strongly depends on the interatomic phase $\phi_z$. For an in-phase configuration (a), the coupling is via the symmetric Dicke state $\ket{+}$, whereas for atoms radiating out-of-phase (b), the cavity couples via the anti-symmetric Dicke state $\ket{-}$.}
\end{figure}

%Cavity quantum electrodynamical systems, such as atoms coupled to a high quality cavity, are usually analysed using linear combinations of the atomic and field states, the dressed states. For higher excitations in the system, however, the dressed state picture and its transitions can become involved \cite{Agarwal:2012}. Here, we will treat the transitions in an undressed framework, in which the emerging phenomena can be explained, while at the same time yielding a very clear picture of the occurring transitions.
We describe the atoms by collective basis states according to the spin-algebra of two two-level systems, i.e., the triplet consisting of the symmetric states $\ket{gg},\ket{+}=(\ket{ge}+\ket{eg})/\sqrt{2},\ket{ee}$, and the anti-symmetric singlet $\ket{-}=(\ket{ge}-\ket{eg})/\sqrt{2}$. The states with a single atomic excitation, i.e., $\ket{\pm}=D_\pm^\dagger \ket{gg}$ are also called symmetric and anti-symmetric Dicke state and are created by the collective Dicke operators $D_\pm^\dagger=(S_1^+ \pm S_2^+)/\sqrt{2}$ \cite{Agarwal:2012}. Together with the $n$-photon Fock state $\ket{n}$ describing the cavity mode, these states fully characterize the occurring transitions. An overview is given in Fig. \ref{fig:transitions} for two different configurations of the atoms.
The transitions seen in Fig. \ref{fig:transitions}(a) show two atoms radiating in-phase with $\phi_z=0 \, (2\pi)$, i.e., they are located at antinodes with an interatomic distance of a multiple of $\lambda_C$. On the contrary, atoms at a distance given by an odd multiple of $\lambda_C/2$ constitute an out-of-phase radiation configuration ($\phi_z=\pi$) where the corresponding transitions are depicted in Fig. \ref{fig:transitions}(b).

The external laser ($\eta$) always pumps symmetric states, i.e., $\ket{+}$ and $\ket{ee}$, regardless of the interatomic phase $\phi_z$. This can be seen by rewriting the pumping Hamiltonian in terms of the collective operators $D_\pm^\dagger$ \cite{Fernandez-Vidal:2007}, which yields $H_L=\hbar \sqrt{2}\eta(D_+^\dagger + D_+)$ and thus gives rise to the transitions 
\begin{equation}
\ket{gg,n}\overset{\eta}{\longrightarrow}\ket{+,n}\overset{\eta}{\longrightarrow}\ket{ee,n} \, .
\end{equation}
The interaction term, however, strongly depends on the interatomic phase $\phi_z$ and couples the atoms to the cavity via $\ket{+}$ or $\ket{-}$. The corresponding Hamiltonian $H_I$ can be rewritten as a function of the interatomic phase reading $H_I(\phi_z)=H_+(\phi_z) + H_-(\phi_z)$, where $H_\pm (\phi_z) = \hbar g_\pm(\phi_z) (a D_\pm^\dagger+a^\dagger D_\pm)$ and $g_\pm(\phi_z)=g[1\pm \cos (\phi_z)] /\sqrt{2}$. 
For the two special cases of $\phi_z=0$ and $\phi_z=\pi$, i.e., in-phase and out-of-phase radiation, respectively, one of the two Dicke states is uncoupled from the atom-cavity interactions.
For atoms radiating in-phase, it obviously holds $g_-(\phi_z=0)=0$ and the anti-symmetric Dicke state $\ket{-}$ is (even almost completely) uncoupled from the dynamics. 
In the case of atoms radiating out-of-phase, however, $g_+(\phi_z=\pi)=0$ and the cavity couples to the atoms via $\ket{-}$.
The atom-cavity interactions can thus be written as 
\begin{subequations}
\begin{eqnarray}
\phi_z &= 0 :  &\ket{ee,n}\overset{g_+}{\longleftrightarrow}\ket{+,n+1}\overset{g_+}{\longleftrightarrow}\ket{gg,n+2} \, , \label{eq:A-C-I-in-phase} \\
\phi_z &= \pi :  &\ket{ee,n}\overset{g_-}{\longleftrightarrow}\ket{-,n+1}\overset{g_-}{\longleftrightarrow}\ket{gg,n+2} \, . \label{eq:A-C-I-out-of-phase}
\end{eqnarray}
\end{subequations}
Though for $\phi_z=\pi$, only the symmetric Dicke state is pumped by the applied coherent field and at the same time only the anti-symmetric Dicke state couples to the cavity, the photon number in the cavity is non-zero due to higher-order processes populating the states $\ket{ee}$ and $\ket{-}$. 
The anti-symmetric Dicke state, for instance, can be populated by spontaneous emission of two fully inverted atoms, but as well be depopulated by a further spontaneous emission process, i.e., 
\begin{equation}
\ket{ee,n} \overset{\gamma}{\longrightarrow} \ket{\pm,n} \overset{\gamma}{\longrightarrow} \ket{gg,n} \, .
\end{equation}
The nonzero output was recently measured \cite{Reimann:2015,Neuzner:2016} and, more recently, it was shown that the emission in this configuration can surprisingly exceed the respective radiation from atoms radiating in phase by far, leading to a hyperradiant emission \cite{Pleinert:2017}.

Spontaneous emission only modifies the atomic state and does not depend on the interatomic phase. It thus takes place for any interatomic phase including $\phi_z=0$ and $\phi_z=\pi$. Note that spontaneous emission is the only non-pseudo-spin-preserving process in the symmetric case ($\phi_z=0$), where both atoms are equally coupled to the cavity. In the asymmetric coupling ($\phi_z=\pi$), spontaneous emission and cavity coupling are non-pseudo-spin-preserving.
Transitions due to cavity decay, on the contrary, do not alter the atomic state but shift the system to a different manifold of atomic states, i.e.,
\begin{equation}
\ket{.,n} \overset{\kappa}{\longrightarrow} \ket{.,n-1} \, .
\end{equation}

\section{Phase Control of Nonclassicality}

We proceed with an investigation of the cavity field with respect to quantum correlations, photon statistics and nonclassicality for different parameters of the system. To work out the $\phi_z$-dependent statistics and quantum signatures of the atom-cavity system, we have to solve the master equation given in Eq. \eqref{eq:master}. The dynamical behavior of the system depends on the specific design of the atom-cavity system and thus on many parameters. In order to keep our discussion fairly general, it becomes inevitable to solve the master equation quite universally by studying the behavior in many different regimes. We accomplished this via numerical techniques \cite{Johansson:2012}, where we solve for the steady-state solution by LU-decomposing the matrices. The numerical convergence was ensured by considering different sizes of the photonic Hilbert space.
%resort to numerical techniques based on QuTiP \cite{Johansson:2012}.}

Whereas the pumping strength $\eta$ and the detunings $\delta, \Delta$ can be easily varied, $g$, $\kappa$, and $\gamma$ are inherent properties of the cavity and the atoms used. In order to focus on the collective behavior of the system, we will mainly work with on-resonance driving and coupling ($\delta=\Delta=0$) throughout this study. As a consequence, we investigate the quantum signatures of the $\phi_z$-dependent statistics for different atom-cavity-laser setups mainly with respect to the freely adjustable pumping strength $\eta$.

The quantum signatures include the study of the normalized second-order correlation function at zero time \cite{Glauber:1963}
\begin{equation}\label{eq:2nd-order-corr-fct}
g^{(2)} \equiv g^{(2)}(0) = \frac{\braket{a^\dagger a^\dagger a a}}{\braket{a^\dagger a}^2} \, ,
\end{equation}
which constitutes a natural witness in the setup and can be obtained by a correlated measurement as depicted in Fig. \ref{fig:cavity_setup}. Depending on the value of $g^{(2)}$, different field statistics of the emitted light can be distinguished: antibunched ($g^{(2)}<1$), coherent ($g^{(2)}=1$), bunched ($1 < g^{(2)} < 2$), thermal ($g^{(2)}=2$) and superbunched ($2 < g^{(2)}$). A value of $g^{(2)}$ smaller than one reveals that the considered photonic field behaves nonclassical.

The nonclassical properties of the photonic field can be further examined by use of the Mandel $Q$ parameter \cite{Mandel:1979}. In terms of moments, $Q$ is given by
\begin{equation}\label{eq:Mandel-Q}
Q = \frac{\braket{ a^\dagger a^\dagger a a} - \braket{ a^\dagger a }^2}{\braket{a^\dagger a}} \, ,
\end{equation}
and can likewise be measured in the HBT setup.
$Q$ reveals the kind of underlying photon statistics. $Q=0$ corresponds to light with Poissonian statistics, i.e., perfectly coherent light, whereas a positive value of $Q$ corresponds to super-Poissonian light, which can be interpreted via fluctuations in the light field and is thus considered as classical. A negative value of $Q$, on the other hand, signalizes sub-Poissonian light, indicating a narrower photon distribution than perfectly coherent light, which is a clear indicator for nonclassicality. From the definitions, it is obvious that $Q$ and $g^{(2)}$ are connected via $Q=\braket{a^\dagger a} (g^{(2)}-1)$. Therefore it holds that $Q<0 \Leftrightarrow g^{(2)}<1$ and hence both quantities indicate the same nonclassicality domain. The system with the smallest $Q$-value, however, does not need to have the smallest $g^{(2)}$-value as well. In this paper, we will study both witnesses: we use $g^{(2)}$ to distinguish domains of different light statistics and will exploit $Q$ to investigate the nonclassical domain further.

\begin{figure*}
	\centering\includegraphics[scale=0.88]{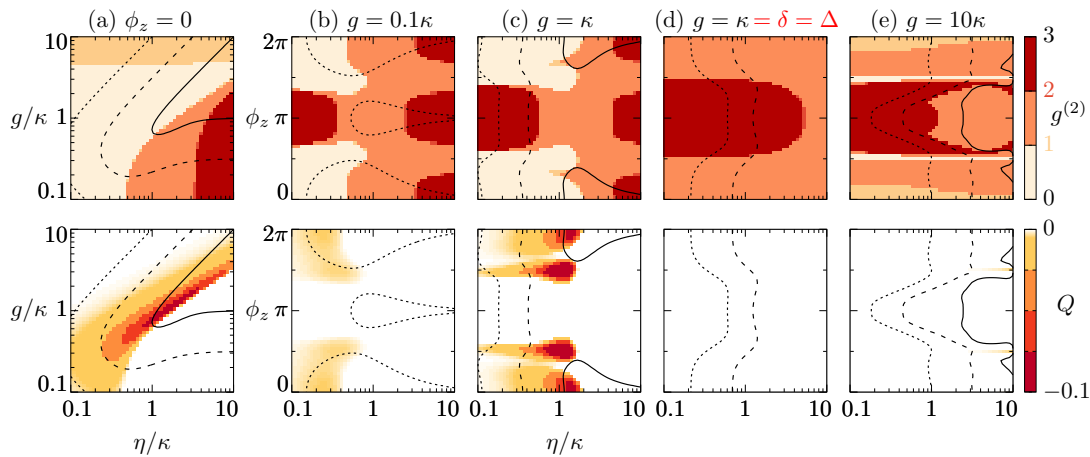}
	\caption{\label{fig:Q_g2a_1} Second-order correlation function $g^{(2)}$ (first row) and Mandel $Q$ parameter (second row) for the system of Fig. \ref{fig:cavity_setup} with $\gamma=\kappa$. In the first row, five different light statistics are distinguished by use of unified colors: from anti-bunched via coherent, bunched and thermal to superbunched. Observe that coherent and thermal light, i.e., $g^{(2)}=1,2$ are also depicted in a unified color in the colorbar. For $Q$, all non-negative values including $Q=0$ are depicted white. In all figures, the dotted, dashed and solid curve indicate the mean photon number $\braket{a^\dagger a}=0.01,0.1,1$, respectively. (a) Atoms radiating in phase. (b),(c),(d),(e) Interatomic phase $\phi_z$-dependent statistics in a bad cavity (b), an intermediate cavity (without (c) and with detuning (d)) and a good cavity (e). The results are discussed in the text.}
\end{figure*}

For in-phase radiating atoms, an overview of the photon statistics can be seen in Fig. \ref{fig:Q_g2a_1}(a) as a function of the couplings to the cavity $g$ and to the external laser $\eta$ reflecting the transitions from bad to good cavities and from low to high pumping, respectively. In these areas, the witnesses reveal a large variety of light statistics. In fact, $g^{(2)}$ in Fig. \ref{fig:Q_g2a_1}(a) covers the entire range from antibunched to superbunched light, while $Q$ can be as small as $-0.1$. 
Including the interatomic phase $\phi_z$ into the discussion yields an even richer picture.  
Plots of the witnesses dependent on $\phi_z$ are shown in Fig. \ref{fig:Q_g2a_1}(b), (c), (d) and (e) for a bad, an intermediate (with and without detuning) and a good cavity, respectively. While at $\phi_z=0 \, (2\pi)$ one is able to recapitulate the in-phase results of Fig. \ref{fig:Q_g2a_1}(a), $\phi_z=\pi$ reveals the out-of-phase radiation of the atoms. We note that at $\phi_z=\pi/2 \, (3\pi/2)$, the second atom is effectively uncoupled from the cavity since $\cos (\pi/2)=0$, yielding a single atom configuration.

In the bad cavity regime $g/\kappa \ll 1$, we can tune the statistic of the light emitted in phase by the pumping strength alone. 
%Low pumping ($\eta \lesssim 0.5\kappa$) yields antibunched nonclassical light, while increasing the pumping can give rise to bunched or even superbunched light ($\eta \gtrsim 3\kappa$). 
In this limit, the dynamics of cavity and atoms can be separated \cite{Bonifacio:1971,Zippilli:2004,Zippilli:2004a}. The cavity field then adiabatically follows the dynamics of the atoms, i.e. $a \propto S^-$. At small $\eta$, every excitation of the atoms is instantly converted into a photon either in the cavity mode ($g_+$) or the sideway modes ($\gamma$). Thus, before the atoms are able to radiate another photon into the cavity mode, the atoms need to get reexcited at first. In this parameter regime, second-order processes are subordinate yielding effectively a two-level system consisting of the excited state $\ket{+}$ and the ground state $\ket{gg}$ similar to the one atom case.  
This - like in the one atom case - leads to antibunched light, see Fig. \ref{fig:Q_g2a_1}(a). At higher $\eta$, however, higher-order processes accompanied by the emission of photon pairs via the transitions given in \eqref{eq:A-C-I-in-phase} are possible and yield bunched or even superbunched light statistics \cite{Temnov:2009}. When $g\lesssim\kappa$, the conditions for coherent and thermal light $g^{(2)}=1,2$ are usually fulfilled only for a very narrow realm. 
In the cavity quantum electrodynamics (CQED) limit $g/\kappa \gg 1$, on the contrary, the cavity field is given by a coherent state for atoms radiating in phase, which can be seen in Fig. \ref{fig:Q_g2a_1}(a) above the threshold $g/\kappa\gtrsim 5$ or in Fig. \ref{fig:Q_g2a_1}(e) at $\phi_z=0$. In this limit, spontaneous emission as well as the pumping only affect the mean photon number. This is in stark contrast to the antibunching observed when driving the cavity instead of the atoms \cite{Brecha:1999} or the antibunched light emitted from one atom, see the thin antibunched region at $\phi_z=\pi/2 \, (3\pi/2)$ in Fig. \ref{fig:Q_g2a_1}(e). In the limit $g/\kappa \rightarrow \infty$, the light field inside a cavity containing a single atom also becomes coherent \cite{Alsing:1992}.
The smallest value of $Q = -0.1$ and therefore the highest nonclassicality is found in an intermediate cavity with $\kappa \approx g \approx \eta \approx \gamma$ and $\phi_z=0$, see Fig. \ref{fig:Q_g2a_1}(a) and (c). In this regime, all occurring effects work against each other leading to nonclassical light. This is similar to the one atom case, see the dark dot at $\phi_z=\pi/2 \, (3\pi/2)$ of Fig. \ref{fig:Q_g2a_1}(c).

\begin{figure}
	\centering \includegraphics[width=0.75\columnwidth]{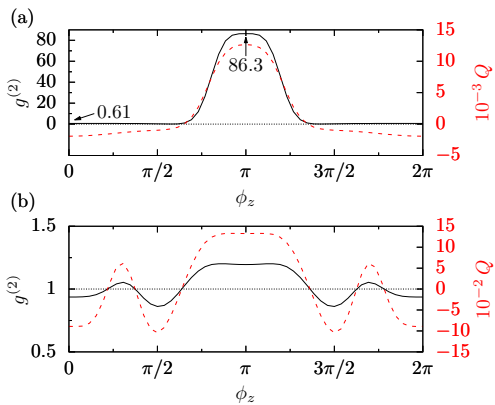}
	\caption{\label{fig:profiles} Profiles of the quantum witnesses of Fig. \ref{fig:Q_g2a_1} at specific values. (a) While at $(g,\eta,\gamma)=(0.1,0.1,1.0)\kappa$, the light field of two in-phase atoms ($\phi_z=0,2\pi$) is antibunched and nonclassical, in the corresponding setup when the second atoms emits a radiation out-of-phase w.r.t. the first atom ($\phi_z=\pi$), a giant photon bunching can be observed. (b) The interatomic phase $\phi_z$ is also a critical factor for nonclassicality. In the presented regime $(g,\eta,\gamma)=(1.0,1.5,1)\kappa$, one can tune the light field from nonclassical to classical by merely moving the second atom.}
\end{figure}

Regardless of the regime, the resulting light field of two atoms radiating out-of-phase ($\phi_z=\pi$) is always bunched or superbunched ($g^{(2)} > 1$), while the nonclassicality vanishes completely ($Q > 0$). In this configuration, the first-order emission into the cavity is forbidden, i.e., $\ket{gg,n}\overset{\eta}{\rightarrow}\ket{+,n} \overset{g}\nrightarrow \ket{gg,n+1}$, leading predominantly to the emission of photon pairs via the antisymmetric Dicke state $\ket{-}$ (which is decoupled from the pumping process), see \eqref{eq:A-C-I-out-of-phase}.  
While in the investigated regimes, the light field of two atoms radiating in phase does not exceed $g^{(2)} = 3$, two atoms radiating out of phase at the same parameters can produce a light field exhibiting a giant photon bunching with $g^{(2)}>80$, see Fig. \ref{fig:profiles}(a). 
A $g^{(2)}$-value of $\approx 60$ in the low pumping regime of two atoms was recently observed by Neuzner et al. \cite{Neuzner:2016}.

Occurrence of detuning radically modifies the situation. In the experimentally realistic scenario $\delta=\Delta$, i.e., an off-resonant driving ($\omega_L \neq \omega_A=\omega_C$), small detuning of the order of $0.1\kappa$ only slightly affects the statistics, while at a detuning $\delta=\Delta\approx \kappa$ (for a good cavity at $\delta=\Delta\approx 10\kappa$) and almost regardless of the pumping, the statistics change to solely bunched light in the case of $\phi_z=0$ and for $\phi_z=\pi$ to even higher $g^{(2)}$-values in the superbunched regime, while the nonclassicality vanishes completely, see Fig. \ref{fig:Q_g2a_1}(d). The same process occurs in bad and good cavities (not shown). Due to the detuning, the excitation probability decreases, and the occurring processes do not compete as much as before. At the same time, the mean photon rate drops.

The resulting photon statistics of the cavity field hence crucially depend on the interatomic phase $\phi_z$ due to the asymmetric coupling of the atoms. By merely moving the atoms (and thus changing the phase), it is possible to tune the output. For a specific and fixed atom-cavity-laser setup, we are thus able to control the photon statistics from antibunching (nonclassicality) to superbunching (classicality) by solely altering the interatomic phase, see Fig. \ref{fig:profiles}(b).

%Observe that one is able to model the transition from one to two atoms with interatomic phase $\varphi$ by scanning $\phi_z=\pi/2 \, (3\pi/2) \rightarrow \varphi$, as at the former value the second atom is effectively decoupled from the cavity. In this way, the modification of the light field when a second atom becomes "visible" to the first one can be investigated. At the same time, the transition from in- to out-of-phase emission, i.e, $\phi_z=\pi$ to $\phi_z=0 \, (2\pi)$, via the one atom configuration can be considered. 

\section{Photon statistics}

In addition to nonclassicality, we also investigate the full photon statistics $p(n)$ of the light field emitted by the setup. Hereby, a noteworthy distribution is the negative binomial distribution (nbd) \cite{Agarwal:1992}, which is able to describe the distribution of diagonal states of classical light. The nbd is defined by
\begin{equation}\label{eq:negbin-distr}
P_{s,p}(n)=\left( \begin{array}{c} s+n-1 \\ n \end{array} \right)  p^s (1-p)^n \, ,
\end{equation}
where the two parameters $s$ and $p$ are restricted to $s\geq 0$ and $0 \leq p \leq 1$. Further, $s$ is restricted to positive integers, but the nbd can be generalized to allow for a continuous $s$ using the Gamma function.
%\begin{equation}\label{eq:negbin-distr-gamma}
%P_{s,p}(n)=\frac{\Gamma (s+n)}{\Gamma (s) \Gamma (n+1)} p^s (1-p)^n \, .
%\end{equation}
%
The parameters of the nbd, $s$ and $p$, can also be related to the moments of the distribution yielding
\begin{subequations}
\begin{eqnarray}
s& = &\frac{\braket{a^\dagger a}^2}{\braket{a^\dagger a^\dagger a a}-\braket{a^\dagger a}^2} \, , \label{eq:s-g2} \\
p& = &\frac{\braket{a^\dagger a}}{\braket{a^\dagger a^\dagger a a}+\braket{a^\dagger a}-\braket{a^\dagger a}^2} \, . \label{eq:p-Q}
\end{eqnarray}
\end{subequations}
%$n=(g^{(2)}-1)^{-1}$ and $p=(1+Q)^{-1}$, 
%where we also incorporated the connection to the already introduced witnesses. 
For $s \rightarrow 1$, it is easy to show that the nbd tends to a thermal distribution (Bose-Einstein distribution) given by
\begin{equation}
P_{th}(n) = \frac{1}{1+\bar{n}}\left( \frac{\bar{n}}{1+\bar{n}} \right)^n \, ,
\end{equation}
with $\bar{n}=\braket{a^\dagger a}=(1-p)/p$.
In this limit, $g^{(2)} \rightarrow 2$ and $0<Q\rightarrow \braket{a^\dagger a}$, which is characteristic for a thermal state \cite{Agarwal:2012}. On the contrary, when $s \rightarrow \infty$ and $p\rightarrow 1$ but $\braket{a^\dagger a}=s(1-p)/p$ stays finite,  $g^{(2)} \rightarrow 1$ and $Q\rightarrow 0$, i.e., the Poissonian statistics of coherent light is recovered. The nbd for diagonal systems can thus be seen as intermediate between coherent and thermal statistics, well-suited to describe (classical) bunched light.

\begin{figure}
	\centering \includegraphics[width=\columnwidth]{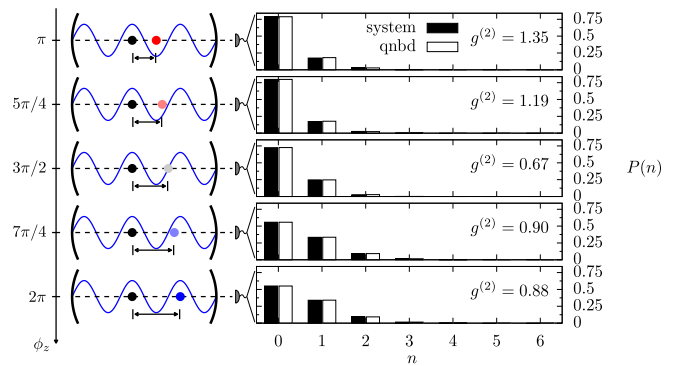}
	\caption{\label{fig:photon-transition} Modification of the photon distribution depending on the interatomic phase $\phi_z$, here at $(\eta,g,\gamma)=(0.7,0.7,1.0)\kappa$. At $\phi_z=\pi$, the second atom emits out of phase with respect to the first atom into the cavity. This leads to bunched light due to the emission of photon pairs via $\ket{-}$, see Fig. \ref{fig:transitions}(b). At $\phi_z=3\pi/2$, the second atom is uncoupled from the cavity yielding effectively a single atom configuration, for which the light field is antibunched. By moving the second atom towards the next antinode of the cavity field to the right, it starts to couple to the cavity again but now radiating in-phase w.r.t. the first atom ($\phi_z=2\pi$). In the entire process, the photon distribution is well described by the qnbd. From the parameters of the qnbd, which here change from $(s,p)=(2.83,0.92)$ to $(s,p)=(-8.7,1.07)$, one can distinguish the classical bunched distribution at $\phi_z=\pi$ and the nonclassical antibunched distribution at $\phi_z=2\pi$.}
\end{figure}

In Fig. \ref{fig:photon-transition}, we consider the aforementioned crossover from bunched to antibunched light when changing from an in- to an out-of-phase configuration with respect to the photon distribution. As expected, the interatomic phase proves to be a critical factor for the resulting photon distribution. 
Moreover, we find that during this classical to nonclassical transition, the photon distribution of the system can be described by a quantum version of the negative binomial distribution (qnbd), which allows for a broader parameter regime as the previously discussed nbd. 
To see this, we use \eqref{eq:s-g2} and \eqref{eq:p-Q} to connect the parameters of the nbd, $s$ and $p$, to the encountered quantum witnesses $g^{(2)}$ and $Q$ yielding $s=(g^{(2)}-1)^{-1}$ and $p=(1+Q)^{-1}$. The usual parameter limitations of the nbd hence directly translate to 
\begin{subequations}
\begin{eqnarray}
s \geq 0  \quad &\Leftrightarrow & \quad g^{(2)}\geq 1 \, , \\
p \leq 1 \quad &\Leftrightarrow & \quad Q\geq 0 \, ,
\end{eqnarray}
\end{subequations}
i.e., bunched light with a \mbox{(super-)}Poissonian statistics. The conventional restrictions of the nbd therefore simply reflect the boundary between classical and nonclassical light. Note that the further restriction $p \geq 0$ leads to the well-known boundary of $Q$ from below, namely $Q\geq-1$, where $Q=-1$ is true for a Fock state \cite{Agarwal:2012}.
When removing these restrictions, i.e., allowing for $s< 0$ and $p>1$, our studies show that the generalized nbd, which we term qnbd, is suited to describe classical and nonclassical states of light. We define the qnbd as the nbd in the classical domain ($P^{(qnbd)}_{s,p}(n)=P_{s,p}(n)$ for $s>0 ,p<1$) and as 
\begin{equation}\label{eq:qnbd}
P^{(qnbd)}_{s,p}(n) =\left\{ 
\begin{array}{ll}
\mathcal{N} P_{s,p}(n) & \mbox{for }  n \leq |s|+1  \, , \\
0 &  \mbox{for } n > |s|+1  \, ,
\end{array}    \right.
 \end{equation}
in the nonclassical domain ($s<0, p>1$).
Naively, one could define the qnbd as the nbd with extended parameter regime. This, however, leads - admittedly only for very nonclassical light fields - to a few inconsistencies such that the so defined qnbd no longer constitutes a valid, i.e., positive and normalized, probability distribution (see appendix for a detailed discussion). 
Hence, we define the qnbd in the nonclassical domain with a cut-off at $n=|s|+1$ and a normalizing factor $\mathcal{N}$ compensating for the cut-off. Note that this cut-off and the normalizing factor are only crucial when investigating very nonclassical light fields (see appendix). Within our studies of the two-atom-cavity system, for instance, we never encountered a regime in which the nontruncated qnbd failed. For the sake of well-definedness, we nevertheless use the definition of \eqref{eq:qnbd}.

%Thereby, we are not restricted to the classical domain as the negative binomial distribution (nbd).

We can also verify the nonclassicality of the qnbd by using Klyshko's criterion \cite{Klyshko:1996} considering neighbouring probabilities:
\begin{equation}\label{eq:klyshko}
\kappa_n = \frac{(n+1)P(n-1)P(n+1)}{nP(n)^2} \, .
\end{equation}
A value of $\kappa_n < 1$ indicates nonclassical behaviour of the corresponding distribution $P(n)$.
For the (q)nbd, we can evaluate \eqref{eq:klyshko} to $\kappa^{(qnbd)}_n=(s+n)/(s+n-1)$. For positive $s$, obviously $\kappa_n > 1$ holds, indicating a classical distribution, i.e., the nbd. For $s<0$, however, $\kappa_n$ is smaller than one as long as $n < |s| + 1$, which coincides with the cutting condition in the definition of the qnbd \eqref{eq:qnbd}. The qnbd can thus be seen as the nonclassical counterpart (or generalization) of the nbd.

In Fig. \ref{fig:photon-transition} at $\phi_z=\pi$, the light field is bunched, while with the same parameters, but at $\phi_z=2\pi$, the light field is antibunched. At all values of $\phi_z$, the qnbd accurately describes this transition. In the classical domain (here at $\phi_z=\pi,5\pi/4$) the qnbd reduces to the nbd with parameters fulfilling the usual restrictions indicating \mbox{(super-)}Poissonian statistics. At, for instance, $\phi_z=3\pi/2$ (single atom) and $\phi_z=2\pi$ (two atoms in phase), the light field is, however, sub-Poissonian, which is reflected by the parameters of the qnbd, which for $\phi_z=2\pi$ yield $(s,p)=(-8.7,1.07)$ clearly violating the usual restrictions of the nbd.

%From the parameters of the qnbd, one could also decide whether the light field is nonclassical, i.e., $s<0,p>1$. Here, at $\phi_z=\pi,5\pi/4,7\pi/4$ the parameters are in the usual range of the nbd indicating \mbox{(super-)}Poissonian statistics, while at $\phi_z=3\pi/2$ (single atom) and $\phi_z=2\pi$ (two atoms in phase), the light field statistics is sub-Poissonian.

\begin{figure}
	\centering \includegraphics[width=\columnwidth]{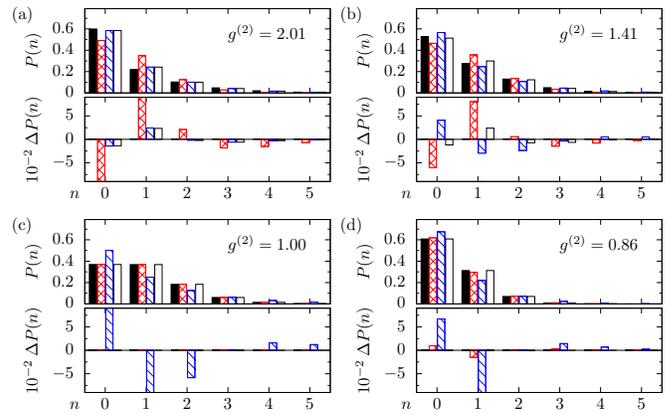}
	\caption{\label{fig:photon-distribution} Comparison of the photon distribution of the system as calculated from \eqref{eq:master} (black, filled) with coherent (red, checked), thermal (blue, lined) and a quantum version of the negative binomial distribution (blank) in different regimes of light statistics: (a) thermal at $(\eta,g)=(4.2,0.8)\kappa$, (b) bunched at $(\eta,g)=(1.5,0.6)\kappa$, (c) coherent at $(\eta,g)=(7.0,7.0)\kappa$ and (d) antibunched at $(\eta,g)=(0.6,0.6)\kappa$ of Fig. \ref{fig:Q_g2a_1}(a). In each subplot, the probability distribution $P(n)$ and the deviations from the photon distribution of the system $\Delta P(n) = P(n)-P_{\rm{system}}(n)$ for up to $n=5$ photons are shown. The parameters of the qnbd are here (a) $(p,s)\approx(0.58,0.99)$, (b) $(p,s)\approx(0.76,2.44)$, (c) $(p,s)\approx(1.00,-540)$, and (d) $p\approx 1.07 >1 $, $s \approx -7.4 <0$. Observe that at the coherent point, the parameters are very sensitive. In (c), $g^{(2)}=0.998$ is actually slightly antibunched leading to the high negative value of $s$.}
\end{figure}

Our study further shows that the photon distribution of the cavity in quite different photon statistics domains can be well described by the (q)nbd. 
Examples of the photon distribution of the system in domains of different statistics are depicted in Fig. \ref{fig:photon-distribution} and compared to coherent distribution, thermal distribution and the quantum version of the negative binomial distribution. In all four domains of Fig. \ref{fig:photon-distribution}, i.e., (a) thermal, (b) bunched, (c) coherent and (d) antibunched, the qnbd is well suited to describe the photon distribution of the system. Particularly in the nonclassical regime of (d), where the parameters of the qnbd are $p\approx 1.07 > 1$ and $s \approx -7.4 <0$, the qnbd fits very well with a maximum deviation from the system of the order of $\Delta P(n) \lesssim 10^{-3}$ and a calculated fidelity of $1$. Note that in all plots, the fidelity is higher than $0.999$.
%In the classical domain, the qnbd reduces to the common negative binomial distribution and can be seen as intermediate between coherent and thermal for diagonal systems.
%
In the superbunched region at high mean photon numbers, however, the qnbd fails to describe the photon distribution within the cavity, which becomes peaked at high values of $n$.

The qnbd constitutes an intriguing tool to study the photon statistics of light fields in quantum optics, as it can be defined in terms of two very well-known nonclassicality measures, i.e., $g^{(2)}$ and $Q$.
In the classical regime, the qnbd reduces to the negative binomial distribution, known since the 90's in the quantum optics community as an intermediate between a thermal and a coherent distribution. In addition, as demonstrated in this article, this distribution can be extended to the nonclassical domain, covering a broad parameter regime.

\section{Conclusion}

We have shown that the photon statistics of the collective radiation of two atoms in a cavity is quite multifaceted. By modifying the interatomic phase only, i.e., merely moving one atom, the resulting cavity light field can be tuned from antibunched to superbunched and from nonclassical to classical. The highest nonclassicality was found for two atoms radiating in-phase into the cavity when all occurring processes were comparable and thus do compete strongly. We were further able to describe the photon distribution in the nonclassical domain by introducing a quantum version of the negative binomial distribution, the parameters of which are directly related to two well-known nonclassicality measures, namely the second-order intensity correlation function at zero time $g^{(2)}(0)$ and the Mandel $Q$ parameter. The experimental conditions that we use have been achieved in the lab recently \cite{Reimann:2015,Neuzner:2016}. We are thus confident that the presented \mbox{$\phi_z$-}dependent photon statistics will stimulate new experimental activities.

\section*{Acknowledgments}
M.-O.P. gratefully acknowledges the hospitality at the Oklahoma State University and financial support by the Studienstiftung des deutschen Volkes. 
The authors gratefully acknowledge funding by the Erlangen Graduate School in Advanced Optical Technologies (SAOT) by the German Research Foundation (DFG) in the framework of the German excellence initiative.
Some of the computing for this project was performed at the OSU High Performance Computing Center at Oklahoma State University supported in part through the National Science Foundation grant OCI–1126330.

\appendix*

\section{Constraints of the quantum negative binomial distribution} \label{sec:appendix-A}

In the main article, we extended the negative binomial distribution $P_{s,p}(n)$ to the nonclassical regime by allowing new parameter ranges, i.e., $s<0$ and $p>1$. In this appendix, we will investigate and discuss the extended probability distribution, the qnbd, in the newly allowed parameter regime of nonclassicality. In order to be a valid probability distribution, the qnbd should be positive (with $0\leq P_{s,p}(n) \leq 1$) and normalized. We will see that in the first definition of the qnbd, i.e., simply extending the nbd without any cut-off, a few inconsistencies can occur, which to some extent, however, can be circumvented.

\begin{figure}
	\centering \includegraphics[width=0.85\columnwidth]{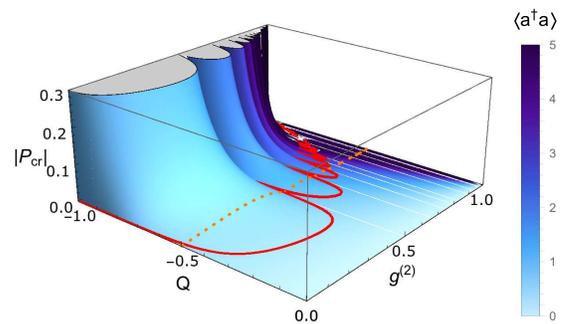}
	\caption{\label{fig:QNBD-restrictions} Absolute value of the critical probability $P_{cr}$ of the nontruncated qnbd in the nonclassical area $(g^{(2)},Q):(0,-1) \rightarrow (1,0)$. The constraints for the nontruncated qnbd are given by the bold red line indicating $|P_{cr}|=0.001$ and by the dashed orange line showing $Q=-0.5$. The colour code reveals the mean photon number $\braket{a^\dagger a}$ which is specified according to $\braket{a^\dagger a}=Q/(g^{(2)}-1)$. Observe that for $g^{(2)}\rightarrow 1$, $\braket{a^\dagger a}$ can become arbitrarily high (the color code here stops at $\braket{a^\dagger a}=5$), and the mean photon number is not specified at the border point to the classical area, i.e., at $(g^{(2)},Q)= (1,0)$. The white lines are due to the divergence of the Gamma function, which can be neglected (see the text for details).}
\end{figure}

The qnbd is, for instance, ill-defined when its parameter $s$ takes the values $s \in \mathbb{Z} \setminus \mathbb{N}_0$, which in terms of experimentally accessible quantities corresponds to $g^{(2)}=(n-1)/n$ with $n\in \mathbb{N}$. These singularities are due to the divergence of the gamma function, but fortunately they are removable since the corresponding limits match each other, i.e., $\lim_{s \to z-} P_{s,p}(n) = \lim_{s \to z+} P_{s,p}(n)$ for $z \in \mathbb{Z} \setminus \mathbb{N}_0$. As a consequence, we are allowed to disregard these singularities.

For the state of zero photons, the qnbd always predicts a positive probability for $n=0$ photons, $P_{s,p}(n=0)=p^s$ as $p$ is positive in the classical as well as nonclassical domain. Using the recursion relation of the qnbd given by
\begin{equation}\label{eq:qnbd-recursion}
\frac{P_{s,p}(n+1)}{P_{s,p}(n)}=\frac{s+n}{n+1}(1-p) \, ,
\end{equation}
we are able to infer a condition for the positivity of the probabilities for higher $n$.
In the classical domain with parameter restrictions $p\leq1$ and $s>0$, it is obvious from \eqref{eq:qnbd-recursion} that the probabilities are positive for all photon numbers $n$ in this domain. The normalization follows directly from the negative binomial distribution.

In the nonclassical regime, however, the parameters $p$ and $s$ are given by $p>1$ and $s<0$. The recursion relation \eqref{eq:qnbd-recursion} thus only predicts a positive $P_{s,p}(n+1)$ as long as $n<|s|$. For $n \geq n_{cr} :=\lfloor|s|\rfloor+2$ (with $\lfloor.\rfloor$ being the floor function), the probabilities can become negative and in fact alternate between negative and positive for large $n$. Yet, depending on the actual photon statistics, the negative probabilities occur for large photon numbers only with neglectable probability. In the example of Fig. \ref{fig:photon-transition}, for instance, the first negative probability is given by $P_{s=-8.7,p=1.07}(n=10)=-1.85\cdot 10^{-14}$.
In order to quantify this observation, we define a critical probability $P_{cr}$ as the first negative probability, i.e., $P_{cr}=P_{s,p}(n_{cr})$. 
Whenever $|P_{cr}|$ is neglectable and the probabilities tend to decrease for $n \geq  n_{cr}$, the nontruncated qnbd is a valid probability distribution in the sense that it is normalized and positive with $0 \leq P_{s,p}(n) \leq 1$. 
Note that the constraints going along with the normalization coincide with the above limits for the critical probability, since the sum of $P_{s,p}(n)$ for all $n \geq n_{cr}$ is approximately zero under the above conditions. 
A condition for decreasing probabilities, i.e., $|P_{s,p}(n+1)|\leq |P_{s,p}(n)|$, can be derived from the recursion relation \eqref{eq:qnbd-recursion}, which for $n \geq n_{cr} > |s|$ can be bounded by
\begin{equation}
\frac{P_{s,p}(n+1)}{P_{s,p}(n)} = \frac{n-|s|}{n+1} (1-p) < (1-p) \, .
\end{equation}
In order to obtain decreasing probabilities the absolute value of the ratio should thus be less than one, leading to $p<2$ or $ Q > -0.5$. 

In Fig. \ref{fig:QNBD-restrictions}, we plot the absolute value of the critical probability as a function of $g^{(2)}$ and $Q$ over the whole nonclassical domain. The color encodes the mean photon number, which is defined for each point according to $\braket{a^\dagger a}=Q/(g^{(2)}-1)$ (except for the classical point at $g^{(2)}=1$, $Q=0$). The thin white lines indicate the divergence of the Gamma function, which can be neglected since the joining limits match each other (see the discussion above). The bold red line and the dashed orange line reflect the two aforementioned constraints of the nontruncated qnbd, namely a non-neglectable critical probability $|P_{cr}|=0.001$ and decreasing probabilities for $Q > -0.5$, respectively. These constraints mark the valid regime of the nontruncated qnbd in the area to the right from the two lines of Fig. \ref{fig:QNBD-restrictions}.

\bibliography{bib}

\end{document}